\documentclass[conference]{IEEEtran}
\IEEEoverridecommandlockouts
\usepackage{cite}
\usepackage{amsmath,amssymb,amsfonts}
\usepackage{algorithmic}
\usepackage{graphicx}
\usepackage{textcomp}
\usepackage{xcolor}
\def\BibTeX{{\rm B\kern-.05em{\sc i\kern-.025em b}\kern-.08em
    T\kern-.1667em\lower.7ex\hbox{E}\kern-.125emX}}

\usepackage{bm}
\usepackage{booktabs}
\usepackage{hyperref}

\def\vs{{\bm{s}}}
\def\vt{{\bm{t}}}

\def\vw{{\bm{w}}}

\def\evt{{t}}

\def\mA{{\bm{A}}}

\def\mE{{\bm{E}}}
\def\mF{{\bm{F}}}
\def\mG{{\bm{G}}}
\def\mH{{\bm{H}}}

\def\mU{{\bm{U}}}
\def\mV{{\bm{V}}}

\def\mX{{\bm{X}}}

\def\mZ{{\bm{Z}}}

\DeclareMathAlphabet{\mathsfit}{\encodingdefault}{\sfdefault}{m}{sl}
\SetMathAlphabet{\mathsfit}{bold}{\encodingdefault}{\sfdefault}{bx}{n}

\def\sZ{{\mathbb{Z}}}

\def\emH{{H}}

\def\emU{{U}}
\def\emV{{V}}

\def\emX{{X}}

\def\emZ{{Z}}

\newcommand{\R}{\mathbb{R}}

\DeclareRobustCommand{\[}{\begin{equation}}
\DeclareRobustCommand{\]}{\end{equation}}

\usepackage[table]{xcolor}

\usepackage{acronym}
\acrodef{MLE}[MLE]{maximum likelihood estimation}
\acrodef{GFlowNet}[GFlowNet]{generative flow network}
\acrodef{ASR}[ASR]{automatic speech recognition}
\acrodef{UASR}[UASR]{unsupervised automatic speech recognition}
\acrodef{G2P}[G2P]{grapheme to phoneme conversion}
\acrodef{DFM}[DFM]{discrete flow-matching}
\acrodef{NAC}[NAC]{neural audio codec}
\acrodef{CTMC}[CTMC]{continuous time Markov chain}
\acrodef{GAN}[GAN]{generative adversarial network}
\acrodef{OOD}[OOD]{out of domain}
\acrodef{ESD}[ESD]{emotional speech dataset}
\acrodef{CDE}[CDE]{controlled differential equation}

\acrodef{MCD}[MCD]{Mel-cepstral distortion}
\acrodef{RMSE}[RMSE]{root mean squared error}
\acrodef{MAE}[MAE]{mean absolute error}

\definecolor{bestblue}{HTML}{BFD7EA}
\definecolor{secondorange}{HTML}{F6D7A7}
\definecolor{worstred}{HTML}{F4B6B6}

\begin{document}

\title{Continuous-Time Acoustic Modelling with Neural Controlled Differential Equations\\
\thanks{This work was supported by the UKRI AI Centre for Doctoral Training in Speech and Language Technologies (SLT) and their Applications funded by UK Research and Innovation [grant number EP/S023062/1]. For the purpose of open access, the author has applied a Creative Commons Attribution (CC BY) licence to any Author Accepted Manuscript version arising.}
}

\author{\IEEEauthorblockN{Mattias Cross}
\IEEEauthorblockA{\textit{Speech and Hearing Group} \\
\textit{University of Sheffield}\\
Sheffield, United Kingdom \\
mcross2@sheffield.ac.uk}
\and
\IEEEauthorblockN{Minghui Zhao}
\IEEEauthorblockA{\textit{Speech and Hearing Group} \\
\textit{University of Sheffield}\\
Sheffield, United Kingdom \\
}
\and
\IEEEauthorblockN{Anton Ragni}
\IEEEauthorblockA{\textit{Speech and Hearing Group} \\
\textit{University of Sheffield}\\
Sheffield, United Kingdom \\
}
}

\maketitle
\begin{abstract}
Text-to-speech (TTS) models commonly address text--speech alignment by expanding phone-level encoder states to frame-level decoder inputs using predicted durations. While this length-regulation step resolves alignment structurally, this use of duration typically changes only where and how often latent states appear, not the values of the states themselves. This paper proposes a continuous-time mechanism for duration-aware acoustic modelling in TTS using neural controlled differential equations (CDEs). We formulate the phone representation as a temporally parameterised control path and use a neural acoustic vector field to produce a continuous-time hidden state whose values evolve with phonetic content and duration-derived timing. The resulting trajectory can be sampled at discrete points and integrated into a standard acoustic decoder pipeline. Objective results contrast CDEs and typical recurrent models. Subjective results suggest that CDE-based models evaluating one phone per step can improve rank-order agreement between synthesised and reference emotion intensity while maintaining comparable emotion-expression quality to a strong baseline. Additional experiments with half-phone step-sizes suggest that temporal resolution changes the trade-off between style tracking and absolute calibration. These results position CDEs as a promising design space for continuous-time and duration-aware style-sensitive TTS.
\end{abstract}
\section{Introduction}
Text-to-speech (TTS) is a core component of many speech-centred technologies, including augmentative and alternative communication (AAC), personalised synthetic voices, conversational agents, and data augmentation pipelines for speech and language systems \cite{leung24_interspeech,nick2023asrtts}. 
Synthesised speech must be intelligible, natural, temporally well structured, and adaptable to different speakers, styles, and interaction requirements.

A central problem in TTS is alignment, one text unit represents many acoustic units, and this alignment is \textit{irregular }in the sense that all phones in an utterance do not have equal duration. Further, the duration and timing of each phone can vary widely across different contexts and speakers. 
A common method to address this is the encoder--decoder architecture, which has seen many formulations \cite{kim2020glowtts,kim21vits,li2025styletts,popov21gradtts,mehta2024matcha}. 
In this framework, an encoder maps a sequence of linguistic labels to a latent representation. 
This representation is then upsampled to match the expected temporal resolution of the acoustic target by repeating each encoder frame to its expected duration. The upsampled representation is then processed by a decoder to predict raw waveforms.
This formulation resolves alignment structurally, but it makes an imperfect assumption: duration changes where and how often latent states appear in the decoder input, but not the values of the latent states themselves.
For example, a lengthened vowel is not simply a short vowel repeated for more frames: its pitch, energy, spectral tilt, and coarticulatory trajectory may evolve across the segment depending on emphasis, phrase position, emotion, and speaking rate. Duration therefore controls not only how long a phone is realised, but also how its acoustic trajectory unfolds. Duration has a key control on how a phone is uttered, and the methods discussed above force the decoder to model this important relationship through frame-wise context and positional cues, rather than introducing duration information into the encoding process.
\begin{figure}[t]
    \centering
    \includegraphics[width=1\linewidth]{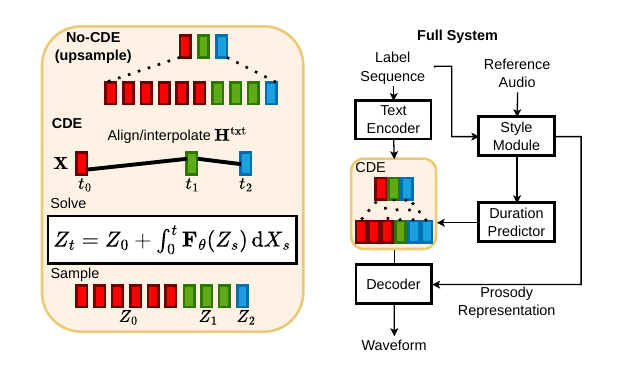}
    \caption{Proposed CDE TTS model. Without a CDE, $\mH^\textnormal{txt}$ is simply aligned by repeating as per the predicted durations, without these durations affecting the values of the decoder input.}
    \label{fig:placeholder}
\end{figure}
A more aligned approach would be to treat TTS as a transformation between a sequence of phone labels observed at irregular time intervals, and a sequence of regularly-sampled acoustic frames. This form of problem is a fundamental topic in the theory of \acp{CDE}, and is the focus of this paper.
Neural CDEs model a dynamically evolving hidden state driven by a control signal, such as a sequence of phone embeddings. The concept of a vector-field being \textit{driven} is associated with rough-path theory \cite{Lyons1998}, where changes in a control signal drive/affect a vector-field. Although they have previously been applied to speech data for classification tasks \cite{ni2025speechemotioncde,kidger2020neuralcde}, their applicability for TTS remains unknown. 
To our knowledge, this work is the first to formulate TTS as a neural acoustic vector-field driven by a continuous-time phonetic latent representation. This formulation exposes temporal resolution as a modelling choice: the phonetic path can be evaluated not only at phone boundaries, but also at intermediate points along the trajectory. This allows duration to influence the values of the latent representation, rather than only determining the number of repeated frames. This method should not be confused with continuous-depth models such as Grad- and Matcha-TTS\cite{popov21gradtts,mehta2024matcha}, which introduce continuous dynamics in the generation process (e.g., diffusion/flow depth) rather than modelling a continuously parameterised phonetic trajectory that explicitly mediates text--time alignment.
To explore the properties of CDEs in TTS, we use a neural CDE for emotion-controlled TTS (Figure \ref{fig:placeholder}). This involves predicting duration and other prosodic features from a reference speech sample, then synthesising an utterance in the same style. Results from subjective evaluation show that a CDE processing one phone per step has a significant Spearman rank correlation of emotion intensity between the synthesised speech and the reference style, whilst maintaining the same quality of emotion expression. 
The properties of continuous-time modelling are further demonstrated by processing a phone-label sequence at a half-phone step-size, a degree of freedom not commonly supported in TTS. This setting offers a less correlated yet more accurate emotion intensity, suggesting that different time resolutions offer different dynamics.
These results suggest that continuous-time models offer a useful mechanism for style-sensitive TTS, and that the temporal resolution of the CDE changes the dynamics of style control. Broadly, this work introduces CDEs as a continuous-time modelling framework for TTS, opening a wider design space in which control paths, interpolation schemes, solver choices, sampling resolutions, and stacked dynamics remain to be explored.
The remainder of this paper is organised as follows:
Section~\ref{sec:method} describes the proposed CDE-based encoder--decoder method.
Section~\ref{sec:experiments} presents the experimental setup.
Section~\ref{sec:results} reports the results and discusses the findings.
Section~\ref{sec:conclusion} concludes the paper. Code is publicly available.\footnote{\url{https://github.com/Mattias421/CDE_StyleTTS.git}}

\section{Method}\label{sec:method}
\subsection{Encoder--Decoder Architecture}
Given a phone-label sequence $\vw \in \sZ^L$ of length $L$, the text encoder $\mE_\theta$ maps it to a C-dimensional phone-level representation,
\[
    \mH^{\textnormal{txt}} = \mE_\theta(\vw) \in \mathbb{R}^{C \times L}.
\]
The latent representation is then upsampled to the length of the reference number of acoustic frames using a product of an alignment matrix $\mH^\textnormal{aco}=\mH^\textnormal{txt}\mA\in\R^{C\times T}$, where $T$ is the target number of acoustic frames and the alignment matrix $\mA \in \{0,1\}^{L \times T}$ is obtained from methods such as Montreal Forced Aligner \cite{mcauliffe17_mfa}, Monotonic Alignment Search ~\cite{kim2020glowtts}, and \ac{ASR} models supporting alignment \cite{graves2006ctc}. This operation is equivalent to repeating each phone embedding according to its predicted or ground-truth duration.  A duration predictor is trained to provide alignments at test time, the acoustic features $\mH^\textnormal{aco}$ are then used by a decoder or decoder + vocoder to produce waveforms.  A duration predictor can either use label feature input, or style features as presented in the next subsection.
\subsection{Style Embeddings}
Encoder-decoder architectures can be enhanced with style embeddings, which allow controllable synthesis guided by the style of a short reference clip \cite{li2023styletts2,xie-etal-2025-towardscontrolspeech}. This involves producing a prosodic text representation $\mH^\textnormal{pros}\in\R^{C\times L}$, using approaches such as a phone-level BERT model adopted by StyleTTS 2\cite{li2023phonebert}. The prosodic representation is then used to condition a diffusion model to predict a style embedding $\vs^{aco}$ conditioned on the reference audio sample. The style and prosodic features are then used to predict pitch, energy, and duration as well as providing additional information to the decoder. Although this method captures style successfully, it introduces a disjoint pipeline where the acoustic embedding $\mH^\textnormal{aco}$ is only affected by duration through the length regulated upsampling, whilst the value of the acoustic embedding remains invariant to duration in every other aspect.

\subsection{Continuous-Time Modelling with CDEs}
Although length regulation provides a form of frame-level alignment, it represents duration only through repetition: the repeated encoder states remain unchanged across frames and therefore do not explicitly encode position-dependent temporal evolution within the alignment.
In particular, two frames may have similar values while originating from phones with different durations or different positions within a phone segment. This weakens the decoder's access to duration-dependent information such as local speaking rate, phone-boundary proximity, and temporal progression within the utterance.

To make this information available to the decoder, we augment the phone embedding with a time channel derived from the alignment $\mA$. Let $\vt = (\evt_1,\ldots,\evt_L)$ denote the frame timestamps induced by the cumulative phone durations, and define the augmented control sequence
\[
    \emX_i = \begin{bmatrix}
        \emH^\textnormal{txt}_i \\ \evt_i
    \end{bmatrix} \in \mathbb{R}^{C+1},
    \qquad i = 1,\ldots,L.
\]
A continuous path $\mX$ is then obtained by interpolating the discrete sequence $\{\emX_i\}_{i=1}^{L}$. Typical interpolations are cubic and linear, this work uses linear and leaves other schemes for future work. Since CDEs are sensitive to change in time, and not absolute time \cite{kidger2021thesis}, the interpolation knots  $\emX_i$ can be placed at regular computational intervals while preserving the temporal information through the values of the time channel itself.

The decoder state is parameterised as the solution of a CDE,
\[
    \emZ_t = \emZ_0 + \int_0^t \mF_\theta(\emZ_s)\,\mathrm{d}\mX_s ,
    \label{eq:cde}
\]
where $\mZ$ is the hidden state, and $\mF_\theta : \R^d\rightarrow\R^{d\times C+1}$ is a neural vector field and $\emZ_0$ is predicted with an initial value network on the observed control sequence $\mX$. In this formulation, the hidden trajectory evolves in response to changes in both the conditioning sequence $\mH^\textnormal{txt}$ and the duration-derived time channel. The CDE therefore provides a continuous-time mechanism for integrating duration information into the decoder, rather than relying only on the repeated or interpolated embeddings produced by length regulation. Note that the formulation of equation \ref{eq:cde} is a Riemann–Stieltjes integral, which is incompatible with black-box ODE solvers, and must be re-written as
\[
    \emZ_t = \emZ_0 + \int_0^t \mF_\theta(\emZ_s)\frac{\mathrm{d}\mX}{\mathrm{d}s}(s)\mathrm{d}s ,
    \label{eq:ode}
\]
which is a matrix-vector product between the vector-field $\mF_\theta(\emZ_s)$ and control derivative $\frac{\mathrm{d}\mX}{\mathrm{d}s}(s)$. For an example of how an ODE solver can be used to sample from the continuous vector field, consider Euler's method
\[
\emZ_{i+1} =\emZ_i +\mF_\theta(\emZ_i)\mathrm{d}\emX_ih,
\]
where $h$ is the selected step size. The formulation of CDEs and their associated ODE solvers is flexible, with several valid choices of control representation and numerical integration scheme. The experiments presented later in this paper use fixed step-size solvers. A step-size $h=1$ corresponds to processing one phone per solver step with a control derivative equal to the difference in the next control frame and the current $\mathrm{d}\emX_{i}h=\emX_{i+1} - \emX_i$. Smaller values of $h$ increase the temporal resolution of the CDE evaluation at the cost of increasing the total number of function evaluations required to solve the CDE. For example, a step size of $h=0.5$ introduces an intermediate solver evaluation within each unit phone interval. Since timing information is incorporated into the control path $\mX$, the vector field $\mF_\theta$ can respond to irregular changes in time, even when it is evaluated in a regular manner.

The resulting hidden state $\mZ$ can either be sampled at the frame-level directly or at the phone-level then aligned as $\mH^\textnormal{aco}=\mZ\mA$, compatible with the decoder during both training and synthesis. Backpropagation through neural CDEs is commonly performed using the adjoint state method \cite{NEURIPS2018_69386f6b}, which can reduce memory requirements by avoiding the storage of the full forward trajectory $\mZ$, making continuous-time modules attractive relative to recurrent and attention-based alternatives for long sequences due to $O(\frac{1}{h}L+d(C+1))$ memory complexity. Further details on training neural differential equations are provided in \cite{NEURIPS2018_69386f6b} and \cite{kidger2021thesis}.

Beyond these training-time memory considerations, neural differential equations offer a unique design space. In particular, each CDE layer produces a continuous hidden trajectory rather than a single static representation. This trajectory can itself be treated as a control signal for a subsequent CDE layer, enabling continuous conditioning modules to be stacked efficiently. The next section describes how this can be achieved.

\subsection{Efficient Stacking of CDE Layers}
The CDE formulation permits multiple continuous-time conditioning layers to be stacked without explicitly materialising and re-interpolating intermediate trajectories \cite{jhin2021attncde,kidger2021thesis}. For example, consider two CDE layers in which the hidden path of the first layer, $\mZ_t$, acts as the control for the second layer:
\[
    \mathrm{d}\mU_t = \mG_\phi(\emU_t)\,\mathrm{d}\mZ_t,
    \qquad
    \mathrm{d}\mZ_t = \mF_\theta(\emZ_t)\,\mathrm{d}\mX_t .
\]
By the chain rule for CDEs, the stacked dynamics can be expressed directly with respect to the control path $\mX_t$:
\[
    \mathrm{d}\mU_t
    =
    \mG_\phi(\emU_t) \mF_\theta(\emZ_t)\,\mathrm{d}\mX_t .
\]
Equivalently, defining the augmented state
\[
    \emV_t =
    \begin{bmatrix}
        \emU_t \\
        \emZ_t
    \end{bmatrix},
\]
the two layers may be integrated jointly as
\[
    \mathrm{d}\mV_t
    =
    \mH_{\theta,\phi}(\emV_t)\,\mathrm{d}\mX_t,
    \qquad
    \mH_{\theta,\phi}(\emV_t)
    =
    \begin{bmatrix}
        \mG_\phi(\emU_t) \mF_\theta(\emZ_t) \\
        \mF_\theta(\emZ_t)
    \end{bmatrix}.
\]
This avoids a separate first-layer CDE solver, interpolation of the sampled states $\mZ$, and a second CDE solver driven by the reconstructed path. Instead, both layers are propagated in a single augmented solve over the same control. The main practical restriction is that the stacked layers share the same numerical solver and integration grid; when different solvers, tolerances, or output resolutions are required, a two-stage interpolation approach remains more flexible.

In the proposed model, this construction allows a hierarchy of CDE-based conditioning blocks to operate on the same duration-augmented control $\mX_t$. The lower layer learns a continuous representation of the aligned phone-duration trajectory, while the upper layer integrates this representation into a higher-level conditioning state.

\section{Experiments}\label{sec:experiments}
\subsection{Data}
\textbf{LJSpeech:} A well-known 24 hour set of audiobook-style narration spoken by a single U.S. speaker\cite{ljspeech17}. Experiments use the same split as \cite{shen2018tacotron2}. Despite its popularity, LJSpeech is prosodically neutral, which may not require the duration-variant features that CDEs provide.  \\
\textbf{Emotional speech dataset:} We follow the original work presented in the first version of StyleTTS\cite{li2025styletts} by using the \ac{ESD} \cite{zhou2021emospeechdata} to measure emotional speech synthesis quality. The English subset of ESD comprises 350 sentences spoken by 10 speakers in 5 emotions (sad, happy, angry, neutral, surprised) totalling 10 hours. A subset of 30 mins per speaker was selected as fine-tuning data.  A random  subset held out for validation and testing containing 100 samples each, split equally across speakers and emotion. For subjective tests, the test split is reduced to 50 samples which are also split equally. The validation and test sets also contain an accompanying reference split which provides each evaluation sample with an utterance from the same speaker and emotion for style prediction. This dataset is challenging because emotional speech relies on subtle prosodic cues, such as dynamic energy and pitch control, which are perceptually salient \cite{larrouy2025emoprosody} but only partially captured by TTS training losses. The use of a 5 hour subset further challenges models to capture nuanced patterns from limited data.

\subsection{Training a CDE within Matcha-TTS}
Recall that Matcha-TTS is an encoder-decoder architecture without style embeddings \cite{mehta2024matcha}. It uses a neural ODE decoder which offers continuous computational depth. To observe how CDEs interact within TTS pipelines, we set up a CDE that uses the Matcha encoder as a control path and Matcha decoder as a readout layer from hidden state to Mel-spectrogram. Only the CDE is trainable, acting as a bridge between text and acoustic representation. This provides a practical test of whether a CDE can be inserted into an existing TTS pipeline without disrupting the pretrained encoder-decoder mapping.

\subsection{Emotional TTS Guided by Reference Style}
The core challenge of this paper involves synthesising speech in the same style as a reference sample, with attention to emotion intensity. This system largely follows Figure \ref{fig:placeholder}, where an encoder produces a control path for a CDE, which is then sampled and then transformed to a waveform by a decoder. The non-CDE modules are initialised by training StyleTTS 2 on LibriTTS (multi-speaker audiobook narration data, neutral emotion) \cite{li2023styletts2}. StyleTTS 2 introduces speech language model (SLM) training, where an adversarial objective is used with WavLM features \cite{chen2022wavlm}, an effective yet computationally demanding process. Several memory-efficient tools are used such as mixed precision, max acoustic length of 175/800 frames, batch size of 6 without SLM adversarial training, and a batch size of 4 with SLM training. The model is trained for a total of 5 epochs, with style diffusion being trained from the second epoch and SLM from the fourth. All modules are trainable. This approach allows finetuning on an NVIDIA A100/H100 GPU with less than 70 GB of video memory. For baseline models we use StyleTTS 2 checkpoints before and after finetuning. This setup uses an ASR-based text aligner. For a strong baseline unrelated to StyleTTS, we consider F5 TTS, which has demonstrated strong zero-shot high-quality synthesis \cite{chen-etal-2024-f5tts}. Studies with autoregressive models are left as future work. Neural CDEs can also be interpreted as continuous-time recurrent models \cite{kidger2020neuralcde}. To examine whether the observed behaviour arises from sequential capacity rather than continuous dynamics, we conduct an experiment comparing several CDE configurations with an LSTM. Both models are trained with the \textit{duration-augmented control path} $\mX$ provided as input.

\subsection{Implementation Details}
First, the initial hidden state $\emZ_0$ is predicted with an LSTM which takes the control sequence in reverse, and compresses the number of channels.  This encodes the full context of the control into the initial state with the initial control state (first phone) $\emX_0$ being observed last, $\emZ_0=\textnormal{LSTM}(\emX_{L,\dots,0})$, providing a strong bias to the CDE. To reduce variance of the control variable $\mX$, we normalise the time channel by the total sum of phone durations for the given utterance. The majority of CDE implementations \cite{kidger2020neuralcde} parametrise the vector-field $\mF_\theta$ with a multi-layer perceptron and a final hidden dimension $d C$, which is then reshaped to form a matrix shape satisfying the ODE in equation \ref{eq:ode}. When the number of control channels is 512 with a hidden dimension of 128, the final linear layer produces a vector of 65536 from 128, a prohibitive expansion. An alternative approach proposed in this paper is to use a U-Net \cite{ronneberger2015unet} with 1 downsampling, 256 middle, and 512 upsampling channels. The U-Net strides along the hidden dimension $d$ and produces the required matrix to evaluate the CDE (Equation \ref{eq:ode}). The CDE is integrated with the fixed-step 4th order Runge-Kutta method \cite{Runge1895UeberDN,kutta1901beitrag}. Once the hidden sequence has been sampled $\emZ_{0,\dots,L}$, it is then upsampled to 512 channels with another U-Net striding frame-wise. A gated residual connection $\mH^\textnormal{cde}=\mH^\textnormal{txt}+\gamma\mZ$ , with trainable parameter $\gamma$, is used to smoothly integrate the CDE module with the rest of the pretrained model during finetuning. The $\beta_1$ hyperparameter of the Adam optimiser is set to 0.9 for the CDE, and 0 elsewhere. The rest of the setup remains unchanged from a typical Matcha-TTS/StyleTTS2 finetuning configuration \cite{mehta2024matcha,li2023styletts2}. Afterwards, the CDE-processed label embeddings are upsampled, ready for the decoder. Although the reparametrisation invariance theorem \cite{kidger2021thesis} suggests operating in label-space provides a sufficiently strong model, an interesting question left unanswered in this work is whether a CDE operating one frame per step rather than one phone per step provides a richer temporal progression at the cost of more ODE solver steps.

\subsection{Evaluation}
The subjective quality of the models is evaluated with emotion intensity mean opinion score (MOS-EI) through a MUSHRA-type set up on the MTurk crowdsourcing platform. A total of 25 participants were issued to listen to a reference audio, then a list of synthetic and ground truth audio permuted in a random order. The listener is only aware which sample belongs to the reference split. Recall that the reference is given to the TTS model to condition style,  but not linguistic content. The participant then rates each clip on a 1-5 integer scale for how strong the emotion is present in the audio e.g. ``how happy is this speech?''. This scheme is inspired by the valence-arousal model of emotion \cite{POSNER_RUSSELL_PETERSON_2005}, where the label, e.g. angry, gives a hint to the target valence-arousal, and the user judges a subjective level of arousal/intensity. The mean opinion of this test reveals how strong different systems express a target emotion, and the correlation of subjective scores to the reference presents how well-calibrated a system is to the intensity of the reference. Overall, each sample was listened to by an average of 10 listeners.
Although MUSHRA is effective at evaluating many systems, listeners pay less attention to subtle effects, motivating comparative MOS (CMOS)\cite{li2023styletts2,ju2024naturalspeech3,minixhofer2024ttsds}. CMOS is an A/B test which displays two samples at a time, and prompts the user to rate which one is better. This work constructs emotion quality CMOS (CMOS-EQ) with a +3/-3 integer scale to represent how better expressed the target emotion is in system B vs system A, with negatives meaning system B is worse than system A. One of the systems will always be the candidate, and the other one of the designated baselines, displayed in random order. In both MOS tests, the listeners are prompted to focus on emotional content. Consistency was checked by duplicating 10 comparisons and secretly swapping the A/B order, all participants self-agreed for at least 5 comparisons. A question arising from these studies is how do listeners perceive unlabelled emotional speech? This is answered by measuring emotion accuracy, which prompts the user with a simple data annotation task given both real and synthetic recordings to identify how well intended emotions are expressed by different systems, and how they are perceived. Each sample is classified by 20 participants. Objective metrics such as \ac{MCD} and log-F0 \ac{RMSE} are used to measure how different design choices affect model outputs. MCD measures the difference between a generated and ground truth Mel spectrogram, with lower values correlating with audio quality/similarity. Log-F0 RMSE measures the error of the pitch contour of the generated signal, lower values correlate with correct intonation, emotion and prosody.

\section{Results and Discussion}\label{sec:results}
\begin{table}
    \centering
    \caption{Effect of training a CDE as a bridge between pre-trained encoder-decoder on LJSpeech.}
    \begin{tabular}{l c c c}
    \toprule
    System & ODE Steps& MCD $\downarrow$ & $\log F_0$ RMSE $\downarrow$ \\
    \midrule
    Matcha-TTS & 10
        & $5.36 \pm 0.52$ & $0.30 \pm 0.08$ \\
    +CDE & 10
        & $5.35 \pm 0.49$ & $0.30 \pm 0.08$ \\
    Matcha-TTS & 5
        & $5.30 \pm 0.52$ & $0.30 \pm 0.07$ \\
    +CDE & 5
        & $5.29 \pm 0.48$ & $0.29 \pm 0.08$ \\
    \bottomrule
    \end{tabular}
    \label{tab:results_matcha}
\end{table}
The results on LJSpeech are displayed in Table \ref{tab:results_matcha}. Given that Matcha-TTS offers continuous computational-depth, we evaluate with both 5 and 10 ODE solver steps. The CDE produces comparable objective scores to the pretrained mapping, with small numerical differences in favour of the CDE conditions. The CDE does not disrupt the encoder--decoder mapping.
\begin{table}[t]
  \centering
  \caption{Objective evaluation results of ESD for the baseline systems and CDE (short and long context).}
  \label{tab:exp4_mcd_f0}
  \begin{tabular}{lcc}
    \toprule
    System & MCD $\downarrow$ & log-F0 RMSE $\downarrow$ \\
    \midrule
    \textit{Short-context (175 frames)} & & \\
    Baseline & $5.30 \pm 0.99$ & \cellcolor{bestblue}$\mathbf{0.40} \pm 0.13$ \\
    LSTM baseline & $5.28 \pm 0.98$ & \cellcolor{secondorange}$0.41 \pm 0.15$ \\
    $h = 0.50$, 1 layer & \cellcolor{secondorange}$\mathbf{5.20} \pm 0.95$ & \cellcolor{bestblue}$\mathbf{0.40} \pm 0.13$ \\
    $h = 0.50$, 2 layers & $5.22 \pm 1.01$ & \cellcolor{bestblue}$\mathbf{0.40} \pm 0.14$ \\
    $h = 1.00$, 1 layer & \cellcolor{bestblue}$\mathbf{5.17} \pm 0.97$ & \cellcolor{worstred}$0.42 \pm 0.14$ \\
    $h = 1.00$, 2 layers & $5.35 \pm 0.96$ & \cellcolor{worstred}$0.42 \pm 0.14$ \\
    $h = 1.00$, 4 layers & \cellcolor{worstred}$5.50 \pm 1.03$ & \cellcolor{secondorange}$0.41 \pm 0.15$ \\
    \bottomrule
    \textit{Long-context (800 frames)} & & \\
    Baseline & $5.18 \pm 0.93$ & \cellcolor{secondorange}$0.40 \pm 0.13$ \\
    F5 baseline & \cellcolor{worstred}$6.24 \pm 1.46$ & \cellcolor{worstred}$0.47 \pm 0.18$ \\
    LibriTTS baseline & $5.57 \pm 0.89$ & $0.41 \pm 0.16$ \\
     $h = 0.50$, 1 layer & $5.20 \pm 0.97$ & \cellcolor{secondorange}$0.40 \pm 0.16$ \\
     $h = 0.50$, 2 layers & $5.15 \pm 0.97$ & \cellcolor{secondorange}$0.40 \pm 0.13$ \\
     $h = 0.50$, 4 layers & $5.26 \pm 0.92$ & \cellcolor{bestblue}$0.39 \pm 0.14$ \\
     $h = 1.00$, 1 layer & \cellcolor{bestblue}$5.04 \pm 0.94$ & $0.41 \pm 0.14$ \\
     $h = 1.00$, 2 layers & \cellcolor{secondorange}$5.07 \pm 0.93$ & \cellcolor{secondorange}$0.40 \pm 0.13$ \\
     $h = 1.00$, 4 layers & $5.24 \pm 0.93$ & \cellcolor{secondorange}$0.40 \pm 0.13$ \\
    \bottomrule
  \end{tabular}
\end{table}

Table \ref{tab:exp4_mcd_f0} shows the objective performance of short-context models (175 frames) over different step-sizes ($h$) and stacked CDE layers. The LSTM and CDE configurations yield different objective behaviour despite receiving the same duration-augmented inputs, indicating that the CDE is not behaving as a direct replacement for the discrete recurrent baseline. The \colorbox{bestblue}{best}, \colorbox{secondorange}{second-best}, and \colorbox{worstred}{worst} results are indicated by \colorbox{bestblue}{blue}, \colorbox{secondorange}{orange}, and \colorbox{worstred}{red}. All models benefitted from longer context. The F5 baseline suffered pacing inaccuracies.

\begin{table}[t]
    \centering
    \caption{Per-emotion reference-style tracking and calibration.}
    \label{tab:emotion_corr_mae}
    \setlength{\tabcolsep}{3pt}
    \resizebox{\columnwidth}{!}{
    \begin{tabular}{lrrrrrr}
    \toprule
    \textbf{System}
    & \textbf{Neu.}
    & \textbf{Ang.}
    & \textbf{Hap.}
    & \textbf{Sad}
    & \textbf{Sur.}
    & \textbf{Macro} \\
    \midrule
    \multicolumn{7}{l}{\textit{Spearman correlation with reference ($\uparrow$) ($p<0.05$ in \textbf{bold})}} \\
    style reference& 1.00 & 1.00 & 1.00 & 1.00 & 1.00 & 1.00 \\

    ground truth
    & 0.29 & -0.15 & 0.33 & 0.29 & 0.14 & 0.18 \\

    F5
    & \cellcolor{bestblue}\textbf{0.66}
    & 0.16
    & 0.26
    & 0.15
    & \cellcolor{worstred}-0.02
    & 0.26 \\

    LibriTTS
    & 0.05
    & 0.12
    & \cellcolor{worstred}0.05
    & 0.48
    & 0.16
    & 0.18 \\

    ESD finetune
    & 0.35
    & \cellcolor{worstred}-0.20
    & 0.15
    & \cellcolor{worstred}-0.30
    & 0.39
    & \cellcolor{worstred}0.08 \\

    h=1.0, 1 layer
    & 0.13
    & \cellcolor{bestblue}\textbf{0.64}
    & \cellcolor{bestblue}\textbf{0.77}
    & \cellcolor{bestblue}\textbf{0.70}
    & 0.19
    & \cellcolor{bestblue}\textbf{0.53} \\

    h=1.0, 2 layers
    & \cellcolor{worstred}-0.16
    & -0.07
    & 0.62
    & 0.30
    & \cellcolor{bestblue}\textbf{0.83}
    & \textbf{0.38} \\

    h=0.5, 1 layer
    & 0.15
    & 0.14
    & 0.25
    & \cellcolor{secondorange}0.55
    & 0.35
    & 0.30 \\

    h=0.5, 2 layers
    & \cellcolor{secondorange}\textbf{0.64}
    & \cellcolor{secondorange}0.20
    & \cellcolor{secondorange}\textbf{0.64}
    & 0.18
    & \cellcolor{secondorange}0.40
    & \cellcolor{secondorange}\textbf{0.43} \\

    \midrule
    \multicolumn{7}{l}{\textit{Mean absolute rating error from reference ($\downarrow$)}} \\
    \addlinespace[2pt]

    style reference& 0.000 & 0.000 & 0.000 & 0.000 & 0.000 & 0.000 \\

    ground truth
    & 0.227 & 0.499 & 0.292 & 0.447 & 0.268 & 0.347 \\

    F5
    & \cellcolor{worstred}0.473
    & 0.429
    & \cellcolor{secondorange}0.255
    & \cellcolor{worstred}0.521
    & \cellcolor{worstred}0.576
    & \cellcolor{worstred}0.451 \\

    LibriTTS
    & 0.305
    & \cellcolor{bestblue}0.343
    & 0.289
    & 0.289
    & 0.572
    & 0.360 \\

    ESD finetune
    & 0.281
    & 0.482
    & 0.288
    & 0.396
    & \cellcolor{secondorange}0.418
    & 0.373 \\
    h=1.0, 1 layer
    & 0.298
    & 0.426
    & \cellcolor{bestblue}0.214
    & \cellcolor{secondorange}0.225
    & 0.543
    & \cellcolor{secondorange}0.341 \\

    h=1.0, 2 layers
    & 0.311
    & \cellcolor{secondorange}0.405
    & 0.262
    & 0.331
    & 0.555
    & 0.373 \\

    h=0.5, 1 layer
    & \cellcolor{secondorange}0.246
    & 0.419
    & 0.355
    & \cellcolor{bestblue}0.157
    & \cellcolor{bestblue}0.403
    & \cellcolor{bestblue}0.316 \\

    h=0.5, 2 layers
    & \cellcolor{bestblue}0.157
    & \cellcolor{worstred}0.538
    & \cellcolor{worstred}0.356
    & 0.310
    & 0.517
    & 0.376 \\
  \midrule
    \multicolumn{7}{l}{\textit{Absolute emotion intensity rating (MOS-EI)}} \\
  style reference& 3.59 & 3.54 & 3.49 & 3.43 & 3.62 & 3.53 \\
  ground truth   & 3.35 & 3.30 & 3.45 & 3.19 & 3.65 & 3.39 \\
  F5             & \cellcolor{worstred}3.09 & \cellcolor{bestblue}3.51 & \cellcolor{secondorange}3.31 &
  \cellcolor{worstred}2.96 & 3.10 & \cellcolor{worstred}3.20 \\
  LibriTTS       & \cellcolor{bestblue}3.50 & \cellcolor{secondorange}3.39 & 3.19 & 3.14 &
  \cellcolor{worstred}3.01 & \cellcolor{secondorange}3.24 \\
  ESD finetune  & 3.31 & 3.20 & 3.20 & 3.20 & \cellcolor{bestblue}3.18 & 3.22 \\
  h=1.0, 1 layer & \cellcolor{secondorange}3.43 & \cellcolor{worstred}3.12 & 3.28 & 3.21 & 3.05 & 3.21 \\
  h=1.0, 2 layers & 3.37 & 3.29 & \cellcolor{bestblue}3.33 & 3.23 & 3.05 & \cellcolor{bestblue}3.25 \\
  h=0.5, 1 layer & 3.41 & 3.21 & \cellcolor{worstred}3.18 & \cellcolor{bestblue}3.30 &
  \cellcolor{secondorange}3.17 & \cellcolor{bestblue}3.25 \\
  h=0.5, 2 layers & \cellcolor{secondorange}3.43 & 3.14 & 3.20 & \cellcolor{secondorange}3.26 & 3.08 & 3.22
  \\
    \bottomrule
    \end{tabular}}
    \end{table}
  
Table \ref{tab:emotion_corr_mae} shows the results of the MUSHRA-style listening test. The first section is the Spearman rank correlation between a system's per-sample rating and the reference, with a macro correlation reported as an aggregate summary across emotions (computed by Fisher $z$-transforming the per-emotion correlations, averaging in $z$-space, and transforming back). The second section is the \ac{MAE}, between a system's intensity rating and the reference. The final section is the overall mean rating, where higher numbers indicate a higher intensity of emotion. Recall that maximising emotion intensity does not indicate a better model, as controllable synthesis requires faithfulness to the reference emotion style, which is captured by the correlation and MAE metrics. The results show that no system is the best across all emotions, and all systems tend to under-express emotion relative to the reference, although the degree of under-expression varies across conditions. The ESD finetune baseline correlates well for surprise, is fair for neutral and happy, but negative for anger and sad. The proposed CDE models have different behaviour depending on their configuration. Using $h=1.0$ and a single layer hinders neutral speech but improves calibration in other emotions, with surprise being worse off. The macro correlation of this CDE (0.53) is higher than the ESD finetuned baseline (0.08). The dynamics of adjusting time step and stacking layers is non-monotonic, but it appears that stacking layers benefits correlation for $h=0.5$, albeit a single layer has less MAE from the reference, and grounds further experimentation into the dynamics of continuous-time discretisation.
\begin{table}[t]
  \centering
  \caption{CMOS-EQ results for CDE $h=1$, 1 layer. Positive values indicate preference for CDE. }
  \label{tab:cmos_dt100_l1}
  \setlength{\tabcolsep}{4pt}
  \resizebox{\columnwidth}{!}{%
  \begin{tabular}{lrrrrrr}
  \toprule
  \textbf{Baseline}
  & \textbf{Neutral}
  & \textbf{Angry}
  & \textbf{Happy}
  & \textbf{Sad}
  & \textbf{Surprise}
  & \textbf{Overall} \\
  \midrule
  ESD finetune& \cellcolor{bestblue}$+0.03$
  & \cellcolor{bestblue}$+0.37$
  & \cellcolor{bestblue}$+0.11$
  & \cellcolor{bestblue}$+0.37$
  & \cellcolor{secondorange}$-0.28$
  & \cellcolor{bestblue}$+0.12$ \\

  Ground truth
  & \cellcolor{secondorange}$-0.50$
  & \cellcolor{secondorange}$-0.90$
  & \cellcolor{secondorange}$-1.34$
  & \cellcolor{secondorange}$-0.93$
  & \cellcolor{secondorange}$-2.01$
  & \cellcolor{secondorange}$-1.14$ \\
  \bottomrule
  \end{tabular}%
  }
  \end{table}

Table \ref{tab:cmos_dt100_l1} shows the results of the emotion expression quality comparison for a 1-layer CDE with $h=1.0$. On average listeners preferred the CDE method over non-CDE baseline for anger and sadness, no preference for neutral and happy, and preferred the baseline for surprise. When scores were averaged per sample, the proposed and baseline systems were statistically indistinguishable under a Wilcoxon signed-rank test, indicating that the two systems produced comparable emotion-expression quality at the utterance-level. However, when averaged per listener, the results showed a significant ($p<0.05$) overall preference for the CDE system, suggesting that listeners exhibited a consistent, albeit modest, preference for the proposed system across the evaluation set. In contrast, the CDE is significantly further than ground truth, highlighting room for improvement. Comparing  with Table \ref{tab:emotion_corr_mae}, the results suggest a partial relationship between calibration and perceived expression quality. This trend follows for surprise.  Neutral-style speech contradicts this but can be considered a special case of non-emotion. Another interesting feature is that the CDE has the lowest intensity rating for anger, yet has +0.37 CMOS-EQ. This could be a compromise where listeners prefer a less intense but more appropriately expressed realisation of anger. Finally, the results for emotion accuracy are in Table \ref{tab:emo_acc}. There is a clear difference between how listeners perceived real speech versus synthetic. Unsurprisingly, LibriTTS overfits to neutral speech, yielding high recall. Between the other systems, the baseline is moderately favoured, with the best accuracy being the 2 layer half step CDE on sadness at $0.558$. From a deployment perspective, it seems that the best strategy is to train multiple model configurations and select the optimal model for a given intended style. Together, the results indicate that CDE-based finetuning can improve faithfulness to the reference emotion style and may also improve perceived expressive quality for some emotions, although these benefits remain dependent on emotion category and configuration. 

\begin{table}[t]
  \centering
  \caption{Human emotion classification accuracy $\uparrow$. }
  \label{tab:emo_acc}
  \setlength{\tabcolsep}{4pt}
  \resizebox{\columnwidth}{!}{%
  \begin{tabular}{lrrrrrr}
  \toprule
  \textbf{System}
  & \textbf{Neu.}
  & \textbf{Ang.}
  & \textbf{Hap.}
  & \textbf{Sad}
  & \textbf{Sur.}
  & \textbf{Macro} \\
  \midrule

  style reference& 0.602
  & 0.639
  & 0.620
  & 0.662
  & 0.858
  & 0.676 \\

  ground truth
  & 0.636
  & 0.780
  & 0.591
  & 0.796
  & 0.653
  & 0.691 \\

  LibriTTS
  & \cellcolor{bestblue}0.606
  & \cellcolor{secondorange}0.437
  & \cellcolor{worstred}0.238
  & \cellcolor{worstred}0.398
  & \cellcolor{worstred}0.087
  & \cellcolor{worstred}0.353 \\

  ESD finetune
  & \cellcolor{secondorange}0.537
  & 0.399
  & \cellcolor{bestblue}0.304
  & \cellcolor{secondorange}0.479
  & \cellcolor{secondorange}0.209
  & \cellcolor{bestblue}0.386 \\

  h=1.0, 1 layer
  & 0.478
  & \cellcolor{bestblue}0.492
  & 0.245
  & 0.469
  & 0.175
  & \cellcolor{secondorange}0.372 \\

  h=0.5, 2 layers
  & \cellcolor{worstred}0.422
  & \cellcolor{worstred}0.318
  & \cellcolor{secondorange}0.294
  & \cellcolor{bestblue}0.558
  & \cellcolor{bestblue}0.255
  & 0.369 \\

  \bottomrule
  \end{tabular}%
  }
  \end{table}

\section{Conclusion}\label{sec:conclusion}
This paper introduced neural controlled differential equations (CDEs) as a continuous-time mechanism for duration-aware acoustic modelling in text-to-speech. The proposed method augments phone-level representations with duration-derived timing information and uses a neural acoustic vector field to produce hidden states whose values evolve with phonetic content and temporal structure, rather than using duration only for length regulation. Experiments show that CDEs provide a small benefit on neutral audiobook-style speech, but can improve reference-style tracking in emotional TTS. In particular, a single-layer CDE with $h=1.0$ achieved the strongest macro Spearman correlation with reference emotion-intensity ratings, while CMOS-EQ results indicated comparable utterance-level emotion-expression quality and a modest listener-level preference over the finetuned StyleTTS 2 baseline. The results also show that step size and layer depth affect the trade-off between style tracking, calibration, and perceived emotion quality, suggesting that temporal resolution is a meaningful modelling choice in CDE-based TTS. This work motivates future investigation into alternative interpolation schemes, solver choices, frame-level control paths, and larger-scale evaluation across more diverse speakers, styles, and data resources.

\section{Acknowledgements}
 Thanks to Aaron Fletcher for proof reading.
We acknowledge IT Services at The University of Sheffield for the provision of services for High Performance Computing.
Portions of the research in this paper used the ESD
 Database made available by the HLT lab, National University of Singapore, Singapore.
The writing of this paper and accompanying code was assisted by ChatGPT and Big Pickle for clearer writing, tables, debugging, and paper/code review.

\bibliography{refs}

\end{document}